\documentclass[journal ]{new-aiaa}
\usepackage[utf8]{inputenc}
\usepackage{textcomp}

\usepackage{graphicx}
\usepackage{amsmath}
\usepackage[version=4]{mhchem}
\usepackage{siunitx}
\usepackage{longtable,tabularx}
\begin{document}
\title{Synchronization framework for modeling transition to thermoacoustic instability in laminar combustors}

\author{Yue Weng$^1$, Vishnu R. Unni$^1$\footnote{vunni@eng.ucsd.edu}, R. I. Sujith$^2$, Abhishek Saha$^1$\footnote{asaha@eng.ucsd.edu}}
\affil{$^1$Department of Mechanical and Aerospace Engineering, University of California San Diego, La Jolla, CA 92093, USA}
\affil{$^2$Department of Aerospace Engineering, Indian Institute of Technology Madras, Chennai 600036, India}
\date{Received: date / Accepted: date}

\maketitle
\begin{abstract}
We, herein, present a new model based on the framework of synchronization to describe a thermoacoustic system and capture the multiple bifurcations that such a system undergoes. Instead of applying flame describing function to depict the unsteady heat release rate as the flame's response to acoustic perturbation, the new model considers the acoustic field and the unsteady heat release rate as a pair of nonlinearly coupled damped oscillators. By varying the coupling strength, multiple dynamical behaviors, including limit cycle oscillation, quasi-periodic oscillation, strange nonchaos, and chaos can be captured. Furthermore, the model was able to qualitatively replicate the different behaviors of a laminar thermoacoustic system observed in experiments by Kabiraj et al.~[Chaos 22, 023129
(2012)]. By analyzing the temporal variation of phase difference between heat release rate oscillations and pressure oscillations under different dynamical states, we show that the characteristics of the dynamical states depend on the nature of synchronization between the two signals, which is consistent with previous experimental findings.
\end{abstract}

\section{Introduction}
In many engineering systems used for power generation and propulsion, such as gas turbines, rocket engines and industrial burners, combustion occurs in a confined space. In such systems, the heat release rate fluctuations can influence the acoustic field of the combustor and vice versa. Such a feedback, when positive, can result in the onset of large amplitude acoustic pressure fluctuations known as thermoacoustic instability~\cite{Lieuwen1}, which is detrimental for both rocket engines~\cite{oefelein} and gas turbines~\cite{Lieuwen1}. The resulting violent oscillations reduce the lifetime of a combustion system and can sometimes lead to serious damage to the hardware. Thus, thermoacoustic instability has become a major concern associated with modern combustor design, especially for newly designed lean-premixed-prevaporized (LPP) combustors, where combustion predominantly occurs in fuel lean conditions to reduce emissions~\cite{unsteadycombustionphysics}. 

A criterion for growth of acoustic energy in a confinement as a result of unsteady heat fluctuation was given by Lord Rayleigh~\cite{Rayleigh}, who explained that when acoustic pressure oscillation is in-phase with the fluctuation of heat release rate, energy is added to the acoustic field. This results in a growth of amplitude of acoustic pressure fluctuations in a thermoacoustic system until it is limited by the nonlinear effects, thus resulting in large amplitude limit cycle pressure oscillations~\cite{Rayleigh,poinsot}. Traditionally, the transition to thermoacoustic instability was considered as a Hopf bifurcation~\cite{Lieuwen2, ananthkrishnan2005reduced}, where the system transitions from a state of fixed point to limit cycle oscillations~\cite{strogatz} as the control parameter is varied. In both experimental and theoretical studies, the transition from a stable non-oscillatory operation to a limit cycle oscillation during the onset of thermoacoustic instability has been thoroughly investigated~\cite{Lieuwen2,Torres,STERLING}.

To model such a thermoacoustic system, and to identify the parameter range where an onset of thermoacoustic instability is expected, the response of the heat release rate to the acoustic fluctuations is traditionally expressed as a flame transfer function (FTF) or flame describing function (FDF). FTF is a linear model of the flame defined in the frequency domain~\cite{CANDEL20021}, as shown in Eqn.~(\ref{eq: FTF}). It depicts the flame's response to small perturbations from the flow (such as pressure fluctuations, flow velocity fluctuations) as a function of frequency~\cite{cheung},

\begin{equation}
\label{eq: FTF}
  F(\omega)=\frac{q(\omega)/Q}{u(\omega)/U}=G(\omega)e^{i\varphi(\omega)}
\end{equation}
Here $F(\omega)$ is the FTF, $q$ and $Q$ are the fluctuation and the mean of the heat release rate, $u$ and $U$ are the fluctuation and the mean of the velocity measured at a constant point respectively, and $\omega$ is the angular frequency~\cite{Macquisten}. FTF for a combustion system can be identified from the experiment and thus, both the excited frequency and the linear growth rate of the system can be derived. It is to be noted that FTF is defined in the limits of small perturbation. In conjunction with a model for the acoustic field in the duct, the FTF approach is able to identify the linear growth rate for acoustic oscillations and consequently, define a stability map~\cite{Paschereit}. However, FTF is unable to predict the effects of nonlinearity, such as the amplitude of limit cycle oscillations, and mode switching~\cite{Bellows}.

FDF, on the other hand, originally introduced by Dowling~\cite{dowling_1997, dowling_1999}, extends the framework of FTF to the nonlinear regime. Based on the theories of control system, FDF is defined as a complex harmonic response of the flame under a sinusoidal forcing and hence is a function of both the frequency and the amplitude of the forcing~\cite{Gelb1968MultipleInputDF}.

\begin{equation}
    \label{eq: FDF}
    F(\omega,|u|)=\frac{q(\omega)/Q}{u(\omega)/U}=G(\omega,|u|)e^{i\varphi(\omega,|u|)}
\end{equation}
Here, $F(\omega,|u|)$ is FDF, $\omega$ and $u$ are the frequency and amplitude of the sinusoidal forcing. As a quasi-linear approach, it considers flame as a nonlinear module coupled to a linear acoustic system. Thus the heat release rate fluctuation can be expressed as a nonlinear response to the acoustic perturbation~\cite{dowling_1997}. In addition to excited frequency and linear growth rate, analysis based on FDF can also predict some nonlinear behavior of the thermoacoustic system, such as the amplitude of a limit cycle oscillation at certain frequency~\cite{dowling_1999}. More recently, Noiray et al.~\cite{Noiray} introduced a unified framework of FDF incorporating the effect of delay in the response of heat release to flow fluctuations. This improved the ability of models to better approximate the experimental findings. However, since the acoustic forcing used for estimating FDF and hence the acoustic field is harmonic, the work done by the heat release rate oscillations at higher harmonics for the given forcing averages to zero over one cycle of forcing. Even though this simplifies the description of FDF, it adds a limiting assumption that acoustic field is always harmonic in time~\cite{ARF_Sujith}.

Furthermore, despite its ability to model the weak nonlinear coupling between acoustics and flame response, FDF cannot model the secondary bifurcations present in a thermoacoustic system. Experimental studies show that even laminar thermoacoustic systems with simple geometry can exhibit elaborate nonlinear behaviors. Kabiraj et al.~\cite{Lipika} investigated various bifurcations in a laminar thermoacoustic system and established a route from self-excited periodic oscillation to chaos via the Ruelle-Takens scenario. Kasturi et al.~\cite{Praveen} discovered bursting oscillations and mixed mode oscillation in a laboratory combustor with a matrix flame. Intermittent oscillations were also observed in a thermoacoustic system as the system transitioned between a state of periodic oscillations to flame extinction~\cite{Lipika2}.  Furthermore, Kashinath et al.~\cite{Kashinath} performed numerical experiments on a coupled flame-acoustic model representing a thermoacoustic system, and examined its dynamical behavior. They found that the thermoacoustic oscillation in a ducted premixed flame can transition to chaos via either the period-doubling route or the Ruelle-Takens-Newhouse route. Recently, strange nonchaos is also found in a self-excited laminar thermoacoustic system by Premarj et al~\cite{Premraj}. Strange nonchaotic attractor (SNA) exhibits complex structure in the phase space. However, the corresponding system does not exhibit sensitivity to initial condition as opposed to a chaotic system~\cite{Gopal}.

Recently, Pawar et al.~\cite{Samadhan} performed experiments on a turbulent combustor with a bluff-body stabilized flame and showed that a synchronization of acoustic field and heat release fluctuation results in the onset of thermoacoustic instability. They have also shown that the synchronized and the desynchronized states correspond to thermoacoustic instability and stable condition respectively and an intermittent phase synchronization state exists between these two states ~\cite{Samadhan}. Inspired from this work, Mondal et al.~\cite{Sirshendu} analyzed the synchronous behaviors of pressure and heat release rate during various states of thermoacoustic oscillations reported by Kabiraj et al.~\cite{Lipika}. They found different synchronization states including phase locking, intermittent phase locking, and phase drifting, corresponding to different dynamical states~\cite{Sirshendu}. Recently, synchronization framework was introduced to model the transitions in aeroelastic system by Raaj et al.~\cite{Ashwad}.

In the flame transfer function or flame describing function approach, the flame is forced at various acoustic velocities, and the response of the flame is evaluated, as a function of the forcing frequency~\cite{Noiray}. The flame transfer function approach does not attribute inherent dynamics to the flame; instead the flame responds to the acoustic velocity fluctuations with a gain and a phase. 

However, in reality the flame in itself is a nonlinear oscillator, that has its own dynamics, and needs to be described as a complete oscillator, not just a quasi-steady response to the acoustic field. This nonlinear oscillator can interact nonlinearly with the acoustic oscillator. When the flame has its own dynamics, we can have different synchronization behaviors~\cite{Sirshendu}. We need a synchronization approach, primarily because we have two oscillators. Thus we attempt to model a laminar thermoacoustic system based on the framework of synchronization. We consider the onset of thermoacoustic instability as the result of synchronization between pressure fluctuation and unsteady heat release. In the present work, we are modeling the bifurcations including the secondary bifurcations and the sequence of bifurcations leading to chaos etc. 

In our model, instead of representing the flame using a transfer function or a describing function, we consider the acoustic field and the unsteady flame as a pair of nonlinearly coupled damped oscillators. A bifurcation analysis is performed by varying the coupling strength. We observed that the model is able to qualitatively replicate the dynamics of thermoacoustic oscillations observed in  experiments.

The rest of the manuscript is organized as follows, in Sec.~\ref{model}, we introduce our model based on the synchronization framework, in Sec.~\ref{experiment}, we describe the experiment from Kabiraj et al. Then, in Sec.~\ref{result}, we compare the dynamics observed in the model and the experiment. The results show that both bifurcations and synchronous behaviors observed in the experiment can be successfully captured in the model. 

\section{\label{model}Model Construction}
The objective of this work is to demonstrate that a synchronization based phenomenological model can qualitatively capture the dynamics exhibited by a laminar thermoacoustic system and can reproduce the various bifurcations in the system. In this study, we however, do not attempt to quantitatively reproduce the experimental observations. We introduce a pair of coupled oscillators to model the thermoacoustic system. Note that for brevity, in the rest of the paper, we use $p(t)$ and $q(t)$ to represent pressure fluctuation and heat release rate fluctuation, respectively. Both the acoustic oscillator and the heat release rate oscillator are considered as damped simple harmonic oscillators that are nonlinearly coupled with each other (Eqn.~({\ref{eq: p oscillator}}), Eqn.~({\ref{eq: q oscillator}})). Thus, when decoupled, for a laminar system, the acoustic oscillator and the heat release rate oscillator (flame) act as damped oscillators that exhibit no oscillations, once steady state is reached.

\begin{equation}
\label{eq: p oscillator}
\ddot p(t)+\zeta_1\dot p(t)+{\omega^2}p(t)={C_{pq}}(1-q(t-\tau_2)^2) \cdot\dot p(t)
\end{equation}

\begin{equation}
\label{eq: q oscillator}
\ddot {q}(t)+\zeta_2\dot{ q}(t)+{R_\omega}^2{\omega^2}q(t)={R_cC_{pq}}(p(t-\tau_1)^3-1)
\end{equation}
Here,
\begin{equation}
\tau_1=T_1=\frac{2\pi}{\omega}\times1,~~ \tau_2=T_2=\frac{2\pi}{R_\omega\omega}\times1
\end{equation}
In Eqn.~(\ref{eq: p oscillator}) and Eqn.~(\ref{eq: q oscillator}), $\omega$ is the angular frequency of the acoustic oscillator, and $R_\omega$ is the ratio of angular frequency of the acoustic oscillator and the heat release rate oscillator. 
The RHS of the equations represent the nonlinear coupling between the two oscillators.
We have not optimized the nonlinear terms used for the coupling. Since the main objective of the paper was to show that a laminar thermoacoustic system could be modeled in a framework of synchronization (i. e., as a set of nonlinearly coupled damped oscillators), our focus was to develop a generic model that captured the qualitative behaviors exhibited in experiments. 
Towards this purpose, we manually chose simple nonlinearities for pressure and heat release rate that resulted in a bifurcation route as observed in experiments.
In our system, the coupling between $p$ and $q$ depends on the local admittance of $q$ in the acoustic oscillator. In the experiments, local admittance varies with the location of the flame in the combustor $x_f$ (explained in the next section). $C_{pq}$ (or $C_{qp}$) is the coupling strength from $q$ to $p$ (or $p$ to $q$), and $R_c = C_{qp}/C_{pq}$. In this study, we consider $C_{pq}$ as the bifurcation parameter. $\zeta_1$ and $\zeta_2$ are the damping parameters of the system.

Previous studies have shown that the delay in the coupling between the unsteady heat release rate and the acoustic fluctuation is an important parameter that determines the dynamics of the system~\cite{Noiray}. Hence, we introduce time delay in the coupling terms of the model. 
For generality, we introduce delay in both the response of both heat release rate fluctuations to the acoustic pressure fluctuations and the response of acoustic fluctuations to the heat release
rate fluctuations. This is reasonable since the delay in either of these subsystems would depend on the position of the flame within the acoustic duct. In previous studies this was not adopted since the models contained only a simple oscillator representing the acoustic field~\cite{STERLING,dowling_1997}.
While the choice of precise time scale of the delay requires further investigation, previous research shows that it is of the order of the time period of oscillation~\cite{Noiray},
and thus for simplicity, in this study the two delays $\tau_1$ and $\tau_2$ are considered to be of the order of time period of the oscillators. The system of differential equations is then solved using Matlab's standard delay differential equation solver DDE23, which uses a three-stage, third-order, Runge-Kutta method with variable time steps.
\raggedbottom

Since the objective of the model is to show qualitative equivalence of the dynamics of the coupled oscillator system with the thermoacoustic system, we chose a generic set of parameters that roughly follows the trends observed in the experiment. Parameters used in the model are,
$\omega^2=160$,
$\zeta_1=\zeta_2=1$,
$R_{\omega}=2.5$,
$R_c=0.9$.
Note that the angular frequency $\omega$ and $R_{\omega}$ in experiments depends on the geometry of the flame holder and the combustor. We observe that the bifurcation diagram is largely unaffected with small variation in $R_{\omega}$ as long as it is of the order of 1. Then, in order to match the frequency distribution of the thermoacoustic system used in experiments, the time $t$ in the model is rescaled. Godavarathi et al.~\cite{Vedasri} showed that the coupling between heat release rate and acoustic fluctuation is asymmetric and that the heat release rate has a greater influence on acoustic fluctuations than vice versa. To account for this we choose the ratio of coupling strength, $R_c = 0.9$ (i.e. $<1$). 

In the present work, we manually tune the nonlinearities in the coupling to obtain a bifurcation diagram similar to that observed in experiments. However, in a future study the optimization of nonlinearities and the parameters of the system can be performed more systematically using data driven modeling methods and tools from machine learning.

\raggedbottom
\section{\label{experiment}Experimental Study}
In this study, we validate the developed phenomenological model for a thermoacoustic system, using the data obtained by Kabiraj et al.~\cite{Lipika} in their experimental study of bifurcations in a laboratory scale laminar combustor. 

The experimental setup consists of a vertical glass duct of length $80~cm$ and inner diameter of $56.7~mm$. The upper side of the duct is open, while the lower side is closed. A coaxial copper tube introduces premixed air-fuel mixture to the burner located inside the combustor. The glass duct is connected to a traverse system that is capable of adjusting the vertical position of the glass duct such that the relative location of the burner (and hence the flame location, $x_f$) within the duct can be varied.  Liquefied petroleum gas (LPG) is utilized as the fuel and it is well premixed with air at a constant equivalence ratio of $\phi=0.48$ prior to being introduced to the burner. The flow rate of the premixed gas mixture is fixed during the experiment. The bifurcation characteristics of the system is studied with variation in $x_f$.

At different values of $x_f$, acoustic pressure fluctuations ($p$) and heat release rate fluctuations ($q$) inside the glass duct are simultaneously measured. A pressure transducer (PCB, 103B02) mounted on the glass duct is used to measure the acoustic pressure fluctuation inside the glass duct. The heat release rate is proportional to the reaction rate. Hence, total $CH^*$ chemiluminesence intensity indicating the global reaction rate is used as a measure of global heat release rate in the system. $CH^*$ chemiluminescence is measured using a photomultiplier tube outfitted with a narrow band optical filter of bandwidth of $10~nm$ centered around $431.4~nm$. Further details of the experimental setup can be found in Kabiraj et al.~\cite{Lipika}. 

\section{\label{result}Results and discussions}
\subsection{Bifurcation analysis}

\begin{figure}
\includegraphics[width=\textwidth]{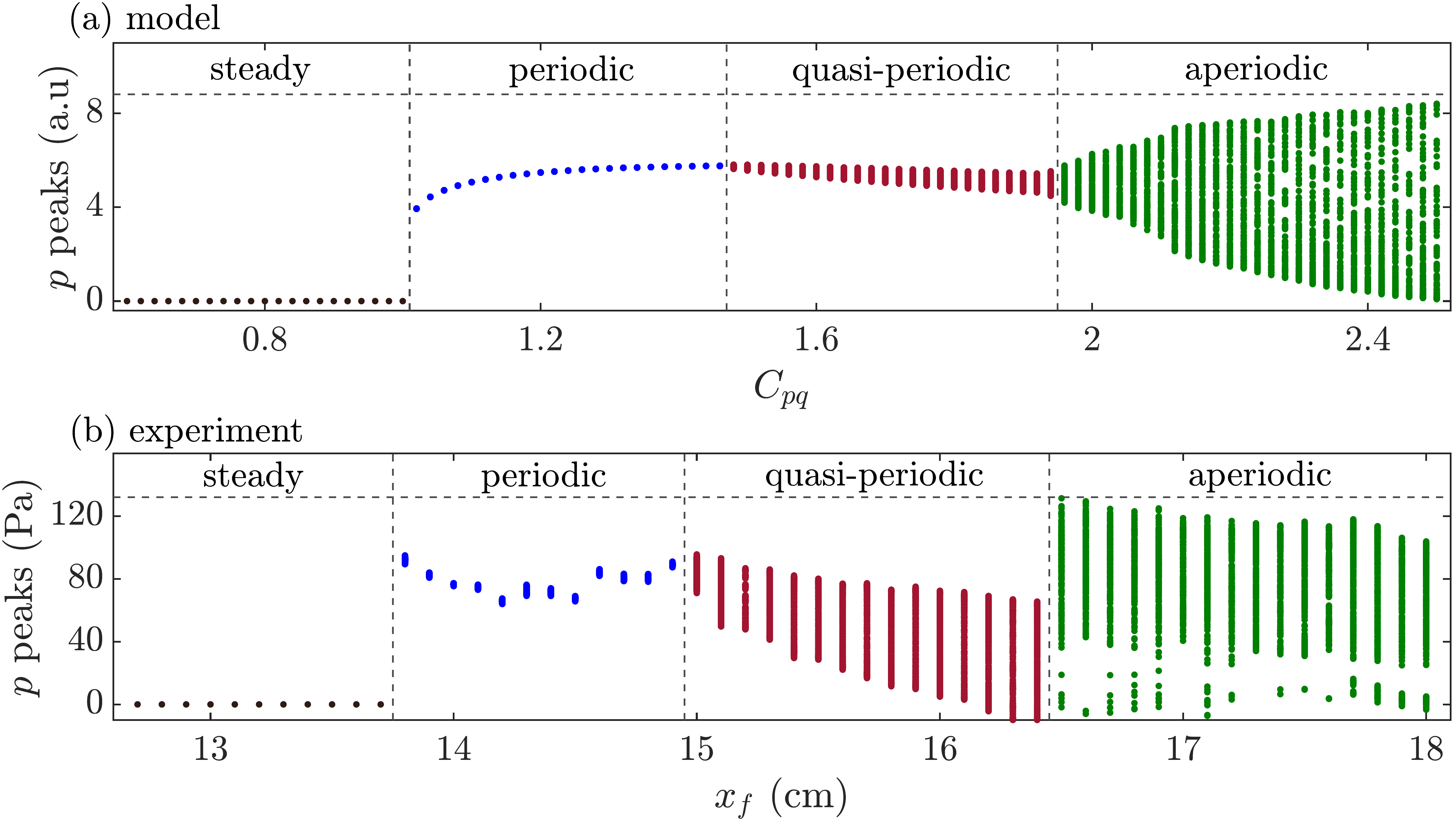}
\caption{\textbf{(a)} A Bifurcation diagram for the model showing the amplitude of local maxima of $p$ as a function of coupling strength $C_{pq}$. \textbf{(b)} A bifurcation diagram from the experiment showing the local maxima of $p$ as a function of $x_f$. In the model, as the coupling strength increases, the system transitions from limit cycle oscillation to aperiodic oscillation through quasi-periodic oscillation, which is consistent with the experiment.}
\label{fig:1}
\end{figure}

\begin{figure}
\includegraphics[width=\textwidth]{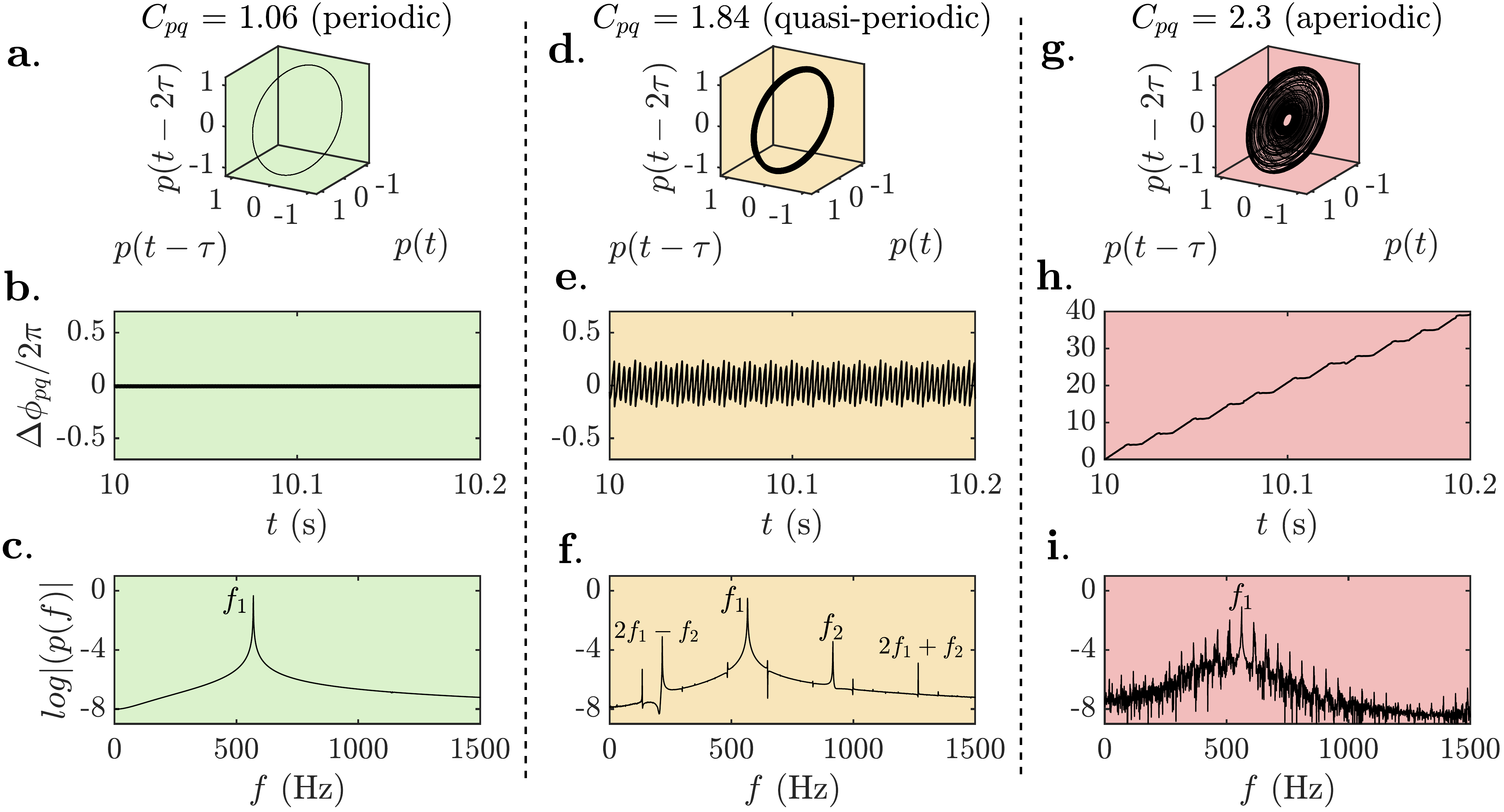}
\caption{The representative phase plot \textbf{(a, d, g)}, phase difference ($\Delta\phi_{pq}$) \textbf{(b, e, h)}, and power spectrum density ($log|(p(f))|$) \textbf{(c, f, i)} are shown from top to bottom for the model. From left to right, periodic ($C_{pq}$~=~1.06,  $f_1~=~570~Hz$), quasi-periodic ($C_{pq}$~=~1.84, $f_1=~566~Hz$, $f_2~=~916~Hz$), and aperiodic ($C_{pq}$~=~2.3, $f_1=~564~Hz$)}
\label{fig:4}
\end{figure}

\begin{figure}
\includegraphics[width=\textwidth]{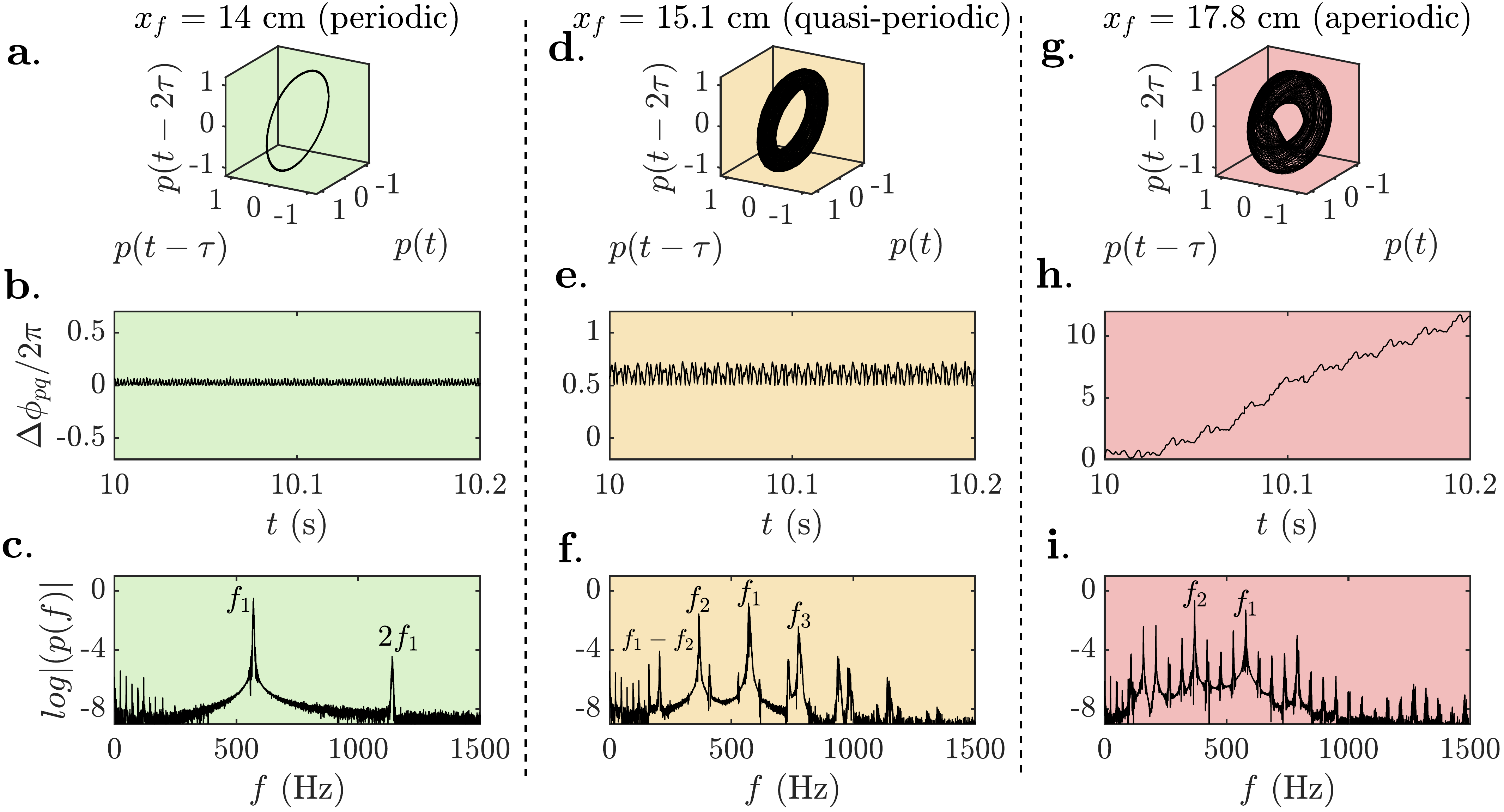}
\caption{The representative phase plot \textbf{(a, d, g)}, phase difference ($\Delta\phi_{pq}$) \textbf{(b, e, h)}, and power spectrum density ($log|(p(f))|$) \textbf{(c, f, i)} are shown from top to bottom for the experiment. From left to right, periodic ($x_f$~=~14~cm, $f_1~=~570~Hz$), quasi-periodic ($x_f$~=~15.1cm, $f_1~=~574~Hz$, $f_2~=~374~Hz$, $f_3~=~747~Hz$), and aperiodic ($x_f$~=~17.8~cm, $f_1~=~579~Hz$, $f_2~=~368~Hz$)} 
\label{fig:5}
\end{figure}

Figure~\ref{fig:1} shows the bifurcation diagrams to track and analyze the qualitative changes in dynamics of the thermoacoustic system with variation of a control parameter. As mentioned before, the coupling depends on the admittance, which varies with flame location ($x_f$) in the experiment, and with the coupling strength $C_{pq}$ in the model. Thus, $C_{pq}$ and $x_f$ are the control parameters in the model (Fig.~\ref{fig:1}(a)) and in experiments (Fig.~\ref{fig:1}(b)), respectively. We plot the local maxima of time series of pressure fluctuations (corresponding to the asymptotic state for a control parameter) as a function of the control parameter. The amplitude distribution of local maxima of time series of a state variable can represent a system's dynamical behavior. For example, a single local maxima represents a constant amplitude, which usually corresponds to limit cycle oscillations. A continuous line with a clear range indicates that the local amplitude varies regularly in a constant range, which may correspond to quasi-periodic oscillations. To further characterize different dynamical natures in the model and the experiment, we also reconstruct the corresponding phase space attractors and analyze the power spectrum of pressure oscillations (Figs.~\ref{fig:4} and \ref{fig:5}).

\begin{figure}
\centering
\includegraphics[width=0.8\textwidth]{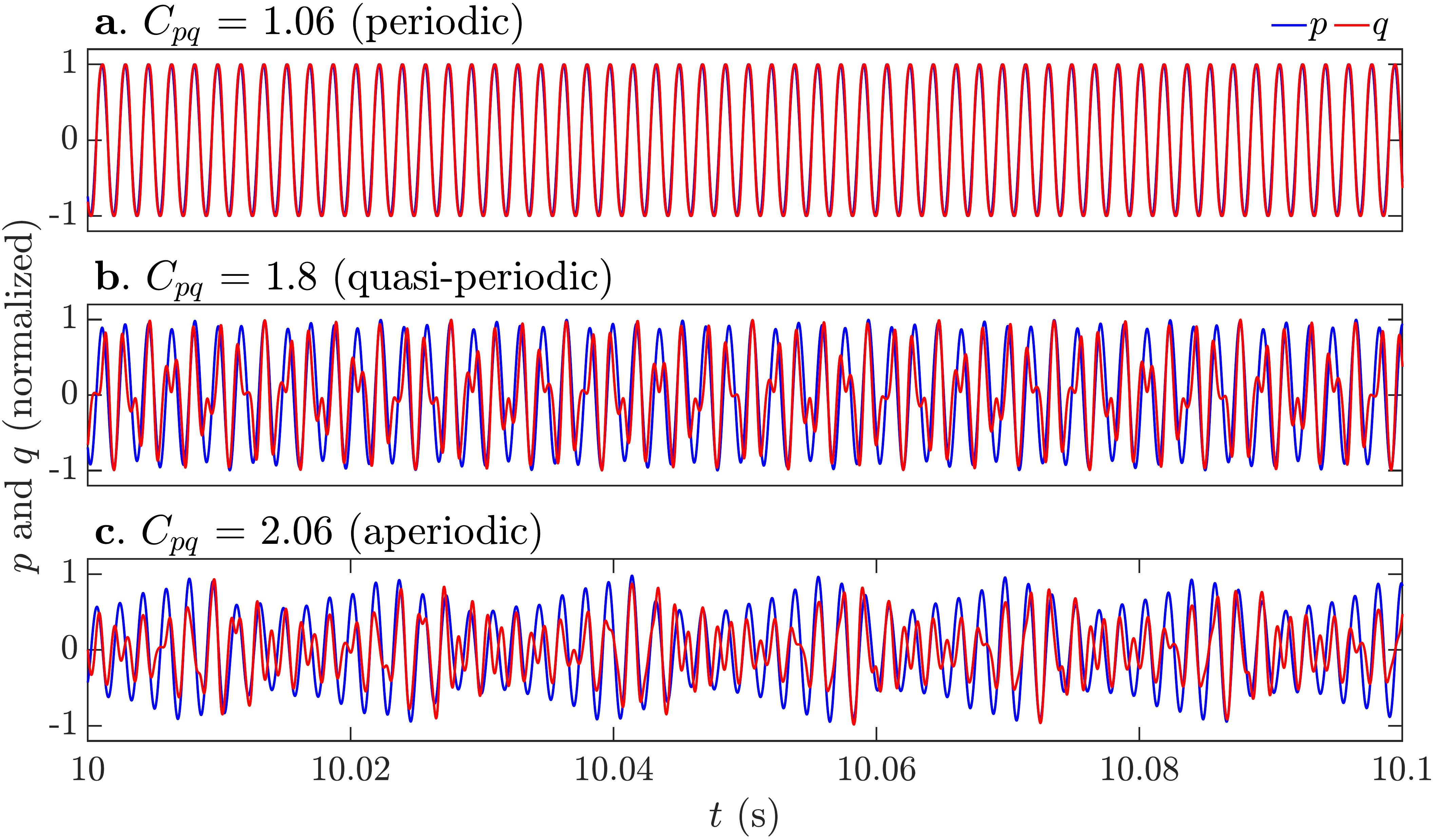}
\caption {The representative normalized time series from the model: from top to bottom periodic ($C_{pq}$~=~1.06), quasi-periodic($C_{pq}$~=~1.8), and aperiodic ($C_{pq}$~=~2.06). During limit cycle oscillation, two time series are mutually synchronized and show a strong periodicity. During quasi-periodic oscillation, irregular periodicity and recurrence are observed and the synchronization is relatively weak. During chaotic oscillation, the two time series are disordered and desynchronized.}
\label{fig:2}
\end{figure}

\begin{figure}
\centering
\includegraphics[width=0.8\textwidth]{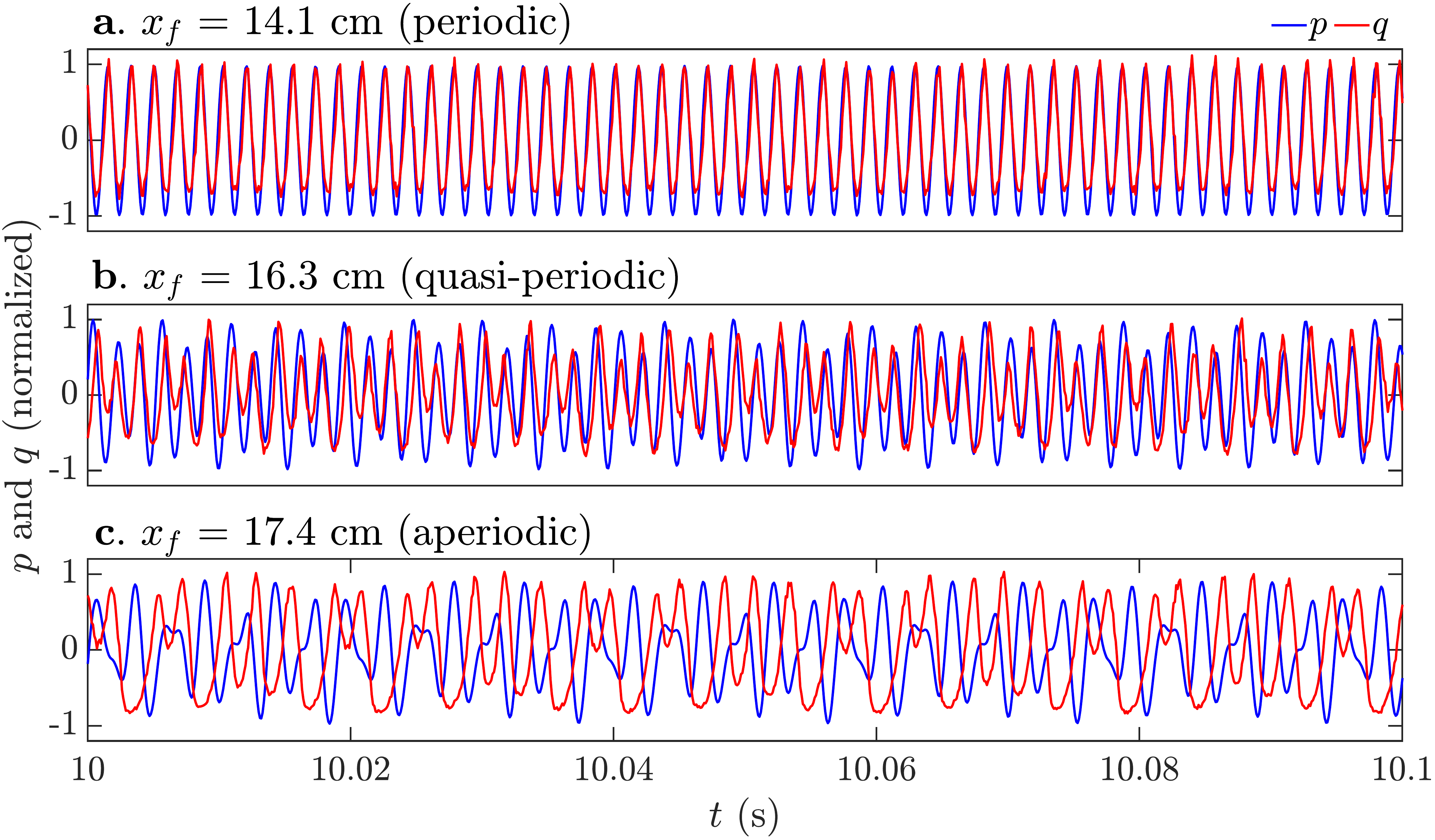}
\caption {The representative normalized time series from the experiments: from top to bottom, periodic ($x_f$~=~14.1~cm), quasi-periodic($x_f$~=~16.3~cm), and aperiodic ($x_f$~=~17.4~cm).}
\label{fig:3}
\end{figure}

As we can see in Fig.~\ref{fig:1}(a), when $C_{pq}$ is small enough, oscillators behave like typical damped oscillators. Since there is one point at zero amplitude for a given coupling strength, the bifurcation diagram, in this region ($0<C_{pq}<1$), consists of a series of fixed zero points (Fig.~\ref{fig:1}(a)). 
As the coupling strength $C_{pq}$ increases, at $C_{pq}=1$ the dynamics change qualitatively and the system exhibits limit cycle oscillations of finite amplitude (Fig.~\ref{fig:2}(a)). This corresponds to a subcritical Hopf bifurcation~\cite{strogatz}. In the corresponding power spectrum, as shown in Fig.~\ref{fig:4}(c), there is only one dominant frequency, which is close to the inherent frequency of the oscillator. In the phase space, the corresponding trajectory is a single closed orbit Fig.\ref{fig:4}(a). At $C_{pq}=1.48$, the second bifurcation occurs (Fig.~\ref{fig:1}(a)), instead of a series of points with constant value, the pattern on the bifurcation diagram now becomes a vertical straight line, which means the local amplitude of the time series in this region is varying continuously in a range (Fig.~\ref{fig:2}(b)). In the corresponding power spectrum, a second frequency peak appears (Fig.~\ref{fig:4}(f)), and the two frequency components are incommensurate with each other. In the phase space Fig.~\ref{fig:4}(d), the closed orbit now becomes a toroidal structure. These features indicate a region of quasi-periodic oscillation, which is caused by a multi-mode interaction. Beyond $C_{pq}=1.96$, with the increase in $C_{pq}$, the time series of $p$ becomes more irregular and disordered (Fig.~\ref{fig:2}(c)), which, in turn, extends the range of peaks in the bifurcation diagram (Fig.~\ref{fig:1}(a)). Moreover, in this range of $C_{pq}$, a broad band of frequencies can be observed in the power spectrum Fig.~\ref{fig:4}(i) and in the phase space, the toroidal structure breaks down to become irregular Fig.~\ref{fig:4}(g). These characteristics indicate that the system is in the aperiodic regime, which can be either chaos or strange nonchaotic attractors (SNAs)~\cite{Premraj, Guan}. In Sec.~\ref{result}.2 we will characterize the dynamics in this regime in detail using the 0-1 test~\cite{Gottwald}.

For the experimental data, we plot the local maxima of $p$ as a function of the flame location $x_f$ Fig.~\ref{fig:1}(b). Comparing the results from the model and the experiments, both dynamical behaviors and the transition between various dynamical states, show that they are qualitatively similar, in that the system transitions from limit cycle oscillation to aperiodic oscillation through a regime of quasi-periodic oscillation.

\subsection{Synchronization behavior of $p$ and $q$}
In the previous section, we demonstrated that our model is able to capture the various dynamics exhibited by the pressure fluctuations. Next, we investigated the capability of model to reproduce the synchronization characteristic of $p$ and $q$ oscillations in the thermoacoustic system. Previously, the synchronization behavior between the pressure fluctuation and heat release fluctuation in the experimental data reported in Kabiraj et al.~\cite{Lipika} was studied by Mondal et al.~\cite{Sirshendu}. Multiple synchronous behaviors were observed from the experiment, including phase locking, intermittent phase locking and phase drifting.

To better understand the interaction between the two coupled oscillators, in Fig.~\ref{fig:4}(b, e, h) we plot the relative phase difference of $p$ and $q$ to distinguish the different synchronization states. We assess the synchronization feature of the model by analysing the variation of the unwrapped instantaneous
phase difference $\Delta\phi_{p,q}(t)$ between the two oscillators.

\begin{equation}
\label{eq: phase difference}
\Delta\phi_{p,q}(t)=\phi_{q}(t)-\phi_{p}(t)
\end{equation}
Here $\phi_{q}(t)$ and $\phi_{p}(t)$ are the phase of $q$ and $p$ respectively.  When the system is in the region of limit cycle oscillation, the relative phase difference between two oscillators is nearly constant with fluctuations near a constant value (Fig.~\ref{fig:4}(b)). In other words, the phase is perfectly locked; this confirms the synchronization behavior and it is consistent with the results from experiment (Fig.~\ref{fig:5}(b)). Although angular frequency in the two oscillators are different, here both signals keep oscillating with the same frequency, which is close to the angular frequency of $p$.

As the system goes to aperiodic regime, intermittent phase locking is observed (Fig.~\ref{fig:4}(h)). For a short period of time, the instantaneous phase differences of the two time series can be locked. However, occasionally, it can jump by multiples of $2\pi$ (Fig.~\ref{fig:4}(h)). Such jumps happen between consecutive phase locked time slots. The intermittent phase locked state will gradually transit to a phase drift state as $C_{pq}$ increases. During the phase drift state, the relative phase difference continuously increases. These observations are also consistent with the findings from experiments (Fig.~\ref{fig:5}(h)). The discussion and analysis of phase difference presented in this sub-section, hence, confirms that the different dynamical states relate to different synchronous behaviors.

\subsection{Distinguishing chaotic state from strange nonchaotic attractors}
Strange nonchaotic behavior can be commonly observed in coupled nonlinear systems~\cite{Nonlinear}. In such a state, a phase space attractor with a fractal geometrical structure similar to chaos usually exists. However, unlike chaotic attractors, the corresponding phase space trajectories do not exhibit sensitivity to initial conditions~\cite{SNA}. Such attractors, which can be identified by the lack of a positive Lyapunov exponent are commonly known as, strange nonchaotic attractors (SNAs). Previous studies have shown that flame response in laminar thermoacoustic systems subjected to either periodic acoustic forcing~\cite{Guan} or self-excitation~\cite{Premraj} can exhibit SNA. Naturally, for our proposed model based on synchronization framework to be successfully used for capturing flame response in thermo-acoustic systems, it is imperative that the model captures SNA. 

Ideally, as the phase space trajectories corresponding to SNAs are not sensitive to initial condition, they can be distinguished from chaos with the lack of positive Lyapunov exponent~\cite{Nonlinear}. However, the data from experiments inevitably contains noise, which restricts the usefulness of the Lyapunov exponent. Hence, we use an alternate method to test chaos, known as 0-1 test~\cite{SNA}. 
In 0-1 test, two translation variables $\mathcal{P}(n)$ and $\mathcal{Q}(n)$ (Eqn.~\ref{eq: var p} \& \ref{eq: var q}) are derived from a one-dimensional time series $\phi(j)$, whose properties are to be investigated. Here $j=1,2,...,N$, $n$ is an index that varies from 1 to $N/10$, and $c$ is a random constant which needs to be chosen in a specific range~\cite{Gottwald}, which we will discuss later.

\begin{equation}
\label{eq: var p}
\mathcal{P}(n)=\sum\limits_{j=1}^n \phi(j) \cos{(jc)}
\end{equation}
\begin{equation}
\label{eq: var q}
\mathcal{Q}(n)=\sum\limits_{j=1}^n \phi(j) \sin{(jc)}
\end{equation}
The behavior of these two translation variables can determine the the dynamical states of the system. More specifically, for a periodic or quasi-periodic time series, $\mathcal{P}(n)$ and $\mathcal{Q}(n)$ are correlated, while for a chaotic time series they are uncorrelated. To further investigate the behavior of $\mathcal{P}(n)$ and $\mathcal{Q}(n)$, a modified mean square displacement can be derived,
\begin{equation}
\label{eq: M}
M(n)=D(n)-V_{osc}(c,n)
\end{equation}
where $D(n)=\displaystyle{\lim_{N\to\infty}\frac{1}{N}\sum\limits_{j=1}^N[\mathcal{P}(j+n)-\mathcal{P}(j)]^2+[\mathcal{Q}(j+n)-\mathcal{Q}(j)]^2}$ is the original mean square displacement and $V_{osc}(c,n)=[\displaystyle{\lim_{N\to\infty} \sum\limits_{j=1}^N}{\phi(j)}]^2\cdot(\frac{1-\cos{nc}}{1-\cos{c}})$ is the correction term to ensure convergence.

The value of $M(n)$ should be bounded within a range around a constant value for a periodic or quasi-periodic time series, and it will linearly increase for a chaotic time series. Subsequently, the asymptotic growth rate of $M(n)$, $K_c$ for a given $c$ is estimated. It can be shown that $K_c$ is the Pearson correlation between vector $\xi=(1, 2, ...,n)$ and $\Delta=(M(1), M(2), ..., M(n))$~\cite{Gottwald2009}. In order to distinguish chaos from other regular dynamics without the effects of resonance, the constant $c$ needs to vary randomly in the range of $\pi/5$ to $4\pi/5$~\cite{Gottwald}. Because the value of $K_c$ depends on $c$, it could attain different values for a given time series. Then the median of these values needs to be calculated to conduct the testing result $K$, which is $K~=~$median$(K_c)$~\cite{Gottwald}. Finally, $K=0$ indicates that the system is nonchaotic while $K=1$ shows the existence of a chaotic attractor~\cite{Gottwald}. To distinguish SNAs from periodic oscillations and chaos, Gopal et al.~\cite{Gopal} showed $c$ should be equal to $\frac{(2\pi)^2}{\omega}$, where $\omega=\frac{\sqrt{5}-1}{2}$.  

\begin{figure}
\includegraphics[width=\textwidth]{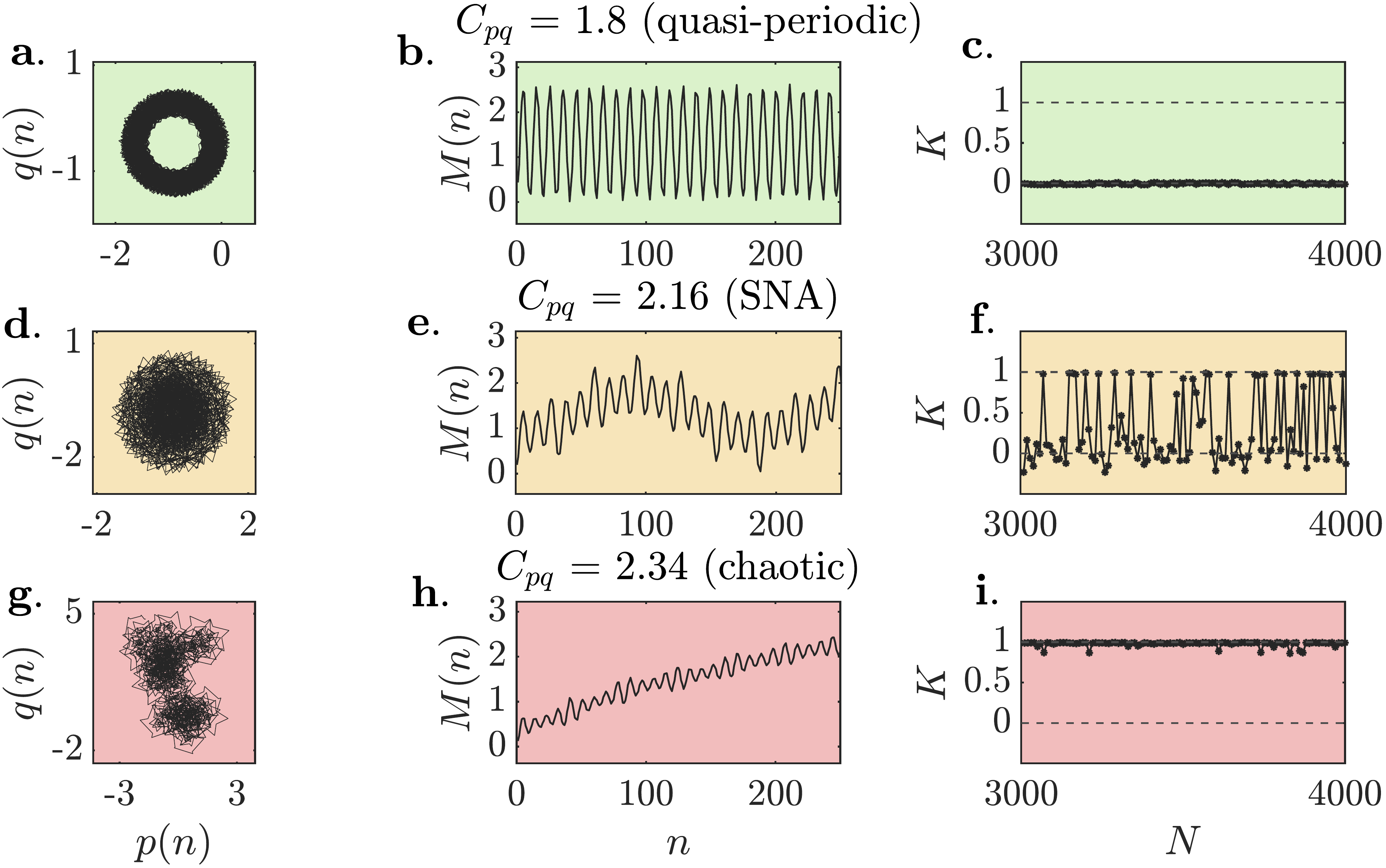}
\caption{The representative diagram on 0-1 test from the model. From left to right: trajectory of translation variables  $\mathcal{P}(n)$ and $\mathcal{Q}(n)$ \textbf{(a, d, g)}, modified mean square displacement $M(n)$ \textbf{(b, e, h)}, the variation of growth rate $K$ \textbf{(c, f, i)}.
Strange nonchaotic attractors are found in the model by applying 0-1 test, which is consistent with the experimental results}
\label{fig:6}
\end{figure}

\begin{figure}
\includegraphics[width=\textwidth]{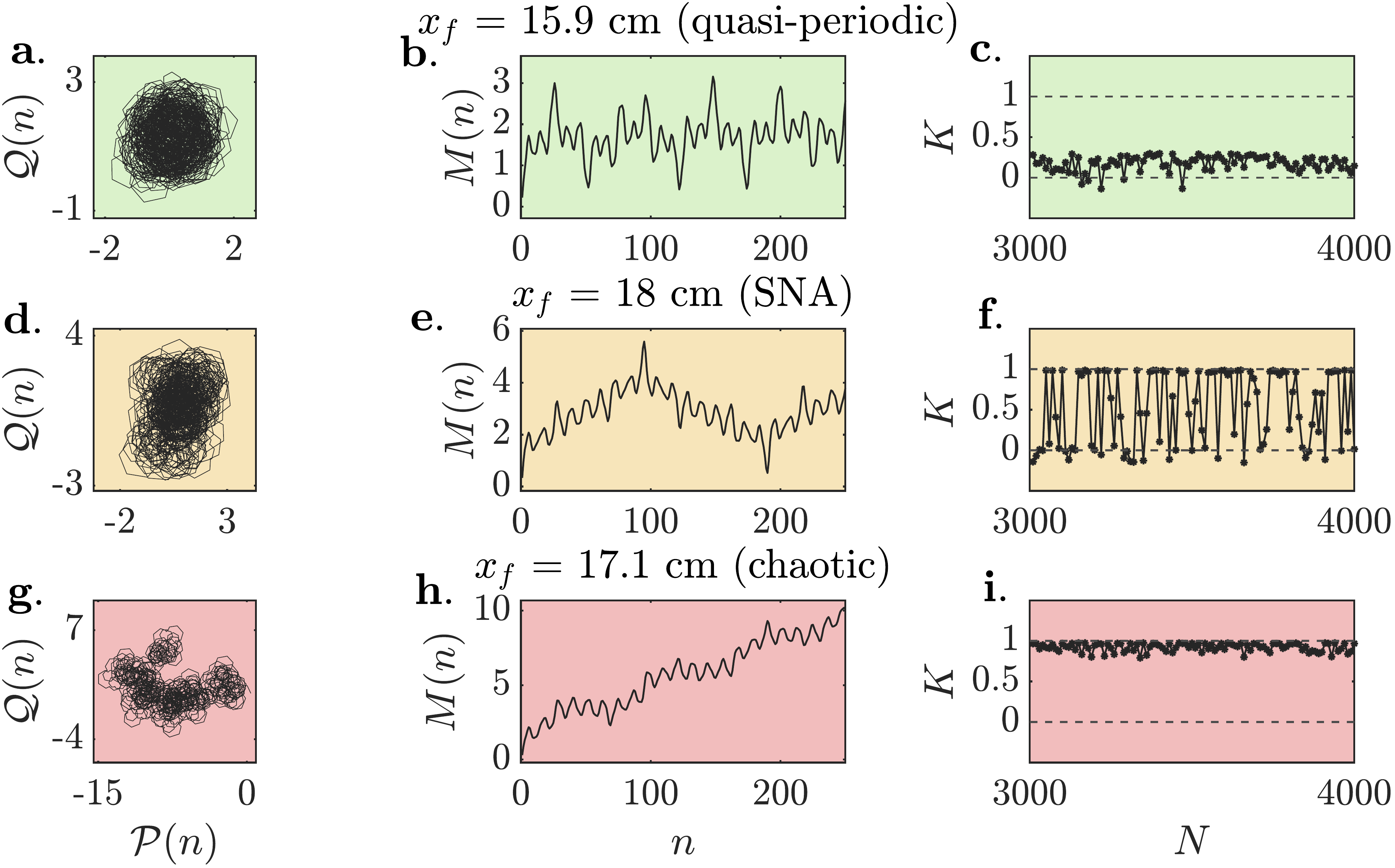}
\caption{The representative diagram on 0-1 test from experiments. From left to right: trajectory of translation variables  $\mathcal{P}(n)$ and $\mathcal{Q}(n)$ \textbf{(a, d, g)}, modified mean square displacement $M(n)$ \textbf{(b, e, h)}, the variation of growth rate $K$ \textbf{(c, f, i)}
}
\label{fig:7}
\end{figure}

Next, we use 0-1 test to various dynamical states captured by the model. The phase portraits of the translation variables  $\mathcal{P}(n)$ versus $\mathcal{Q}(n)$ (Fig.~\ref{fig:6}(a, d, g)), modified mean square displacement $M(n)$ (Fig.~\ref{fig:6}(b, e, h)), and asymptotic growth rate $K$ (Fig.~\ref{fig:6}(c, f, i)) for different dynamics in the model are shown in Fig.~\ref{fig:6}. As we can see, when the system is quasi-periodic, the trajectory of $(\mathcal{P},\mathcal{Q})$ is like a circle (Fig.~\ref{fig:6}(a)), $M(n)$ fluctuates around a constant value (Fig.~\ref{fig:6}(b)), and the asymptotic growth rate $K$ (Fig.~\ref{fig:6}(c)) is always close to zero~\cite{Gopal}. When the thermoacoustic system transitions to chaos, the trajectory of $(\mathcal{P},\mathcal{Q})$ fragments into inordinance (Fig.~\ref{fig:6}(g)) and the modified mean square displacement $M(n)$ is almost linearly increasing (Fig.~\ref{fig:6}(h)), which corresponds to the diffusive behavior of the chaotic attractor, thus $K$ is always close to 1 (Fig.~\ref{fig:6}(i)) ~\cite{Premraj, Gopal}. As for SNA, it can be difficult to distinguish because it has characteristics of both chaotic attractors (strange and complicated structure) and quasi-periodic attractors (does not sensitively rely on initial condition), but it can be recognized from an oscillatory behavior of $M(n)$ (Fig.~\ref{fig:6}(e)). More specifically, $M(n)$ for an SNA for small $n$ may grow linearly. However, for large $n$, $M(n)$ tends to be bounded within a range and oscillates around a constant value \cite{Premraj, Gopal}. Furthermore, the value of $K$ can be lying between 0 and 1 (Fig.~\ref{fig:6}(f)) \cite{Gottwald2004}. Such dynamical behaviors in the model has also been observed in the experimental data (Fig.\ref{fig:7}). Although we are able to detect SNAs in some time series from the model, we note that the range of SNAs is quite narrow in the data from both model and experiment. Thus, currently we are unable to identify a clear boundary for SNA and chaos. We will investigate SNAs predicted by this model in more detail in a future study.

\subsection{Summary and outlook}
In this paper, we introduced a phenomenological model based on the framework of synchronization that captures the dynamical behaviors observed in a laminar thermoacoustic system. Instead of regarding the unsteady heat release rate as a quasistatic response to the acoustic perturbation, we considered the unsteady heat release rate and the acoustic field as a pair of coupled oscillators. By changing the coupling strength, we show that the the model predicts transitions from a state of limit cycle oscillation to chaos through a route of quasi-periodic oscillation, which is consistent with the experimental result observed by Kabiraj et al.~\cite{Lipika}. In addition, strange nonchaotic attractors that appear in the experiment~\cite{Premraj} was also captured in the model. These, in summary, confirm that the proposed synchronization based model using coupled oscillators is able to capture the nonlinear dynamical behaviors of a laminar thermoacoustic system.

The present model still has various limitations. For example, it is hard to analytically determine the expression of the unsteady heat release rate as an oscillator. We may be able to derive it using data-driven optimization methods. Secondly, the model is only tested and compared to laminar thermoacoustic instabilities, while in reality thermoacoustic instabilities usually happen in turbulent combustors. We will consider the effect of turbulence and turbulent combustion in a future work.

\subsection*{\textbf{Acknowledgement}}
We thank Dr. Lipika Kabiraj for sharing the experimental data with us to validate our model. The research at UC San Diego was supported by internal grants. RIS acknowledges the support of ONRG (Grant No. N62909-18-1-2061, contract monitor, Dr. R. Kolar).

\subsection*{\textbf{Compliance with ethical standards}}

\subsection*{\textbf{Conflict of interest}}
The authors declare that they have no conflict of interest.

\newpage
\bibliography{paper}
\end{document}